\documentclass[%
twocolumn,floatfix,
 reprint,
superscriptaddress,
showkeys,
nofootinbib,
amsmath,amssymb,
aps,
pre, 
]{revtex4-2}

\usepackage{graphicx}
\usepackage{xcolor}
\usepackage{amsmath}
\usepackage{amsfonts}
\usepackage{graphicx}
\usepackage{dcolumn}
\usepackage{bm}
\usepackage{comment}
\usepackage{float}
\usepackage{hyperref}

\newcommand {\afb}[1] {{\bf\color {red}[afb : #1]}}

\begin{document}

\title{Latent geometry emerging from network-driven processes}

\author{Andrea Filippo Beretta$^\dagger$}
\affiliation{Dipartimento di Fisica e Astronomia “Galileo Galilei”, Universit\`a degli studi di Padova, via Marzolo 8, 35131 Padova, Italy}

\author{Davide Zanchetta$^\dagger$}
\affiliation{Dipartimento di Fisica e Astronomia “Galileo Galilei”, Universit\`a degli studi di Padova, via Marzolo 8, 35131 Padova, Italy}

\author{Sebastiano Bontorin}
\affiliation{Fondazione Bruno Kessler, Via Sommarive 18, 38123 Trento, Italy}

\author{Manlio De Domenico}
\email{manlio.dedomenico@unipd.it}
\affiliation{Dipartimento di Fisica e Astronomia “Galileo Galilei”, Universit\`a degli studi di Padova, via Marzolo 8, 35131 Padova, Italy}
\affiliation{Istituto Nazionale di Fisica Nucleare, Sez. Padova, Italy}
\affiliation{Padua Center for Network Medicine, University of Padua, Via F. Marzolo 8, 35131 Padova, Italy}

\begin{abstract}
    Understanding network functionality requires integrating structure and dynamics, and emergent latent geometry induced by network-driven processes captures the low-dimensional spaces governing this interplay. Here, we focus on generative-model-based approaches, distinguishing two reconstruction classes: fixed-time methods, which infer geometry at specific temporal scales (e.g., equilibrium), and multi-scale methods, which integrate dynamics across near- and far-from-equilibrium scales. Over the past decade, these models have revealed functional organization in biological, social, and technological networks.
\end{abstract}

\maketitle
\def\thefootnote{$\dagger$}\footnotetext{These authors contributed equally to this work}

\section{Introduction}

The remarkable functionality of complex networks --- from neuronal circuits orchestrating cognitive processes to metabolic pathways sustaining life --- emerges from a delicate interplay between their intricate wiring and the dynamics unfolding upon it~\cite{newman2003structure,boccaletti2006complex,dorogovtsev2008critical,gao2012networks,holme2012temporal,liu2016control,battiston2020networks,rosas2022disentangling,de2023more}. However, while topological information provides insights about a system, for many practical applications where empirical networks change, evolve or adapt, structure alone offers an incomplete picture~\cite{gross2008adaptive,perc2013evolutionary,berner2023adaptive,sole2024fundamental,de2025architecture}. In fact, it is the dynamic patterns of signalling -- such as flows, synchronization, and cascading interactions -- that truly define how information, resources or perturbations propagate and transform over time~\cite{barzel2013universality,hens2019spatiotemporal}. Understanding this intertwined relationship is crucial to unravel the organizational principles that govern complex systems across biology, technology and society, motivating frameworks~\cite{de2016spectral,edler2017mapping,ghavasieh2020statistical,ghavasieh2023generalized,villegas2023laplacian,peixoto2025network,zhang2025coarse} that jointly analyze topology and dynamics to reveal hidden functional architectures~\cite{rosvall2008maps,rosvall2009map,viamontes2011compression,lambiotte2015random,nicolini2017community,ghavasieh2021unraveling,villegas2022laplacian,bovet2022flow,ghavasieh2024diversity}.


The first pioneering studies adopted and developed geometrical methods to address some of the aforementioned challenges, by integrating the structural features of a complex network with geometry~\cite{chung2002average} (see~\cite{boguna2021network} for a review). In this context, the discovery of self-similarity patterns in the structure of complex networks~\cite{song2005self,goh2006skeleton} has been a key step that allowed the development of Renormalization Group techniques for complex networks \cite{radicchi2008complex,rozenfeld2010small}. Subsequently, these methods have received a more complete and profound understanding with the introduction of a hyperbolic geometry framework, in which the network is embedded in a latent hyperbolic space~\cite{serrano2008self,boguna2009navigability,krioukov2010hyperbolic,papadopoulos2012popularity}. Indeed, unlike their spatial counterparts~\cite{barthelemy2011spatial}, complex non-spatial networks do not possess a natural embedding space: accordingly, several approaches have been proposed to embed their structure into a latent geometrical space, each tailored to specific applications and purposes~\cite{goyal2018graph,boguna2021network,baptista2023zoo}. 

\begin{figure*}[!t]
        \centering
        \includegraphics[width=1.0\textwidth]{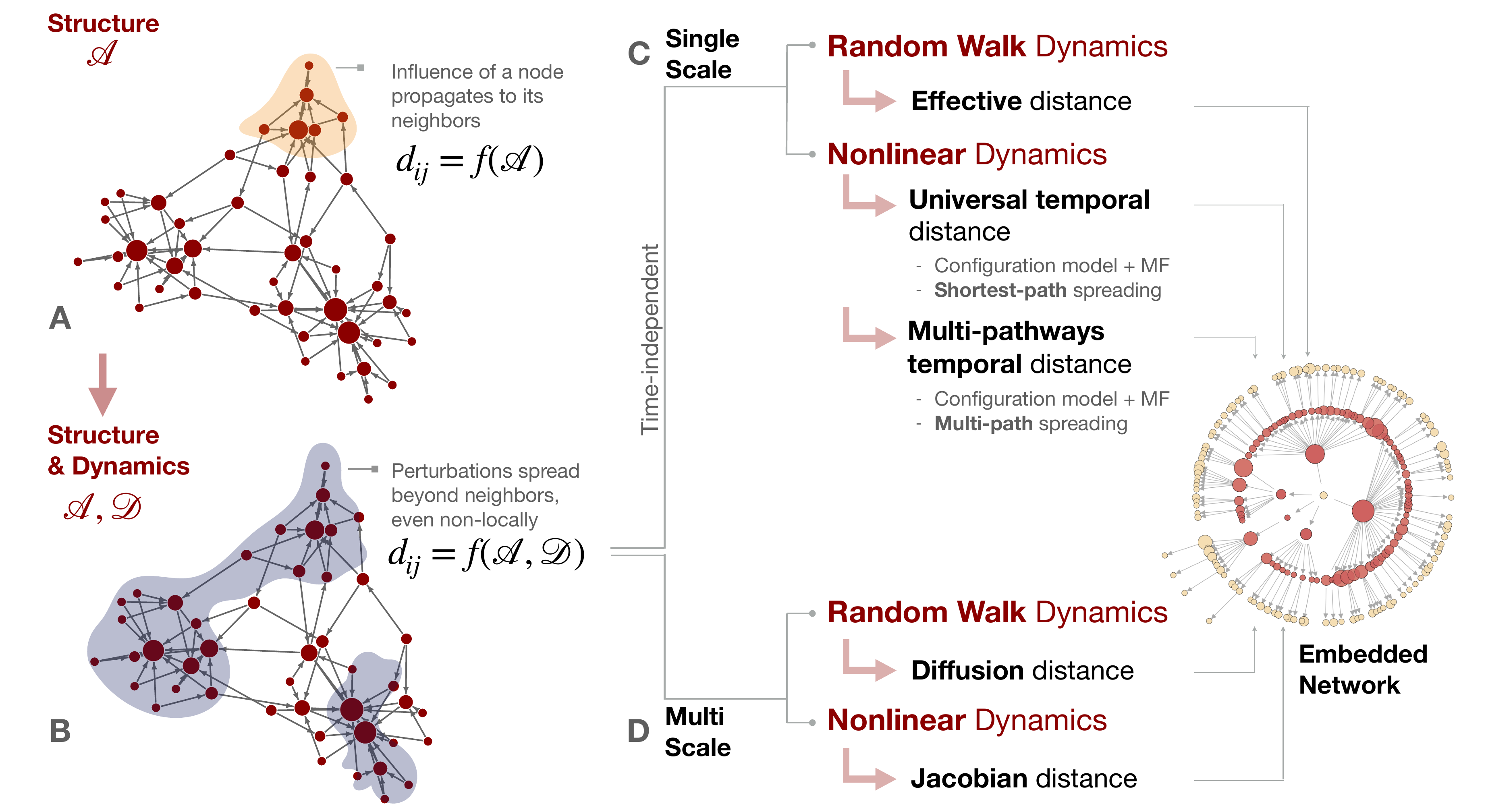}
\caption{\textbf{Functional organization from the interplay between structure and dynamics: latent geometry emerging from network-driven processes}. Latent geometry based on structure (A) defines distances in terms of neighbors properties or structural properties (e.g., shortest paths), while neglecting dynamics. By taking dynamics into account (B), the global properties of signal propagation is uncovered and naturally taken into account for the analysis of network functionality with respect to specific generative models for information exchange among system's units. Two alternative approaches capture complementary information: single-scale distances (C), built upon asymptotic relaxation or effective rates, and multi-scale distances (D), which can be tuned, for instance, to detect mesoscopic network structures by using time as a multi-resolution parameter to probe network functionality.}
    \label{fig:fig1}
\end{figure*}

 While useful and rich of important insights, these structure-based methods are somehow limited by the lack of comprehensive inclusion of dynamical features, which play a crucial role in determining the functionality of empirical complex systems. Therefore, in this work, we overview some promising methods developed to unravel the complex interplay between dynamics and network structures through the lens of latent geometry, induced by network-driven processes. To this aim, we propose a basic yet effective taxonomy to better understand the (dis)similarities among existing approaches: at the highest level, we distinguish between fixed-time scale (e.g., at equilibrium) and multi-scale (e.g., near equilibrium or far from equilibirum) methods (see Fig.~\ref{fig:fig1}).

Firstly, we consider discrete time diffusion processes, where node-node interactions are linear, for which a quasi-metric is obtained \cite{brockmann2013hidden,iannelli2017effective,klamser2024inferring}. We review methods \cite{barzel2013universality,bontorin2023multi} to characterize arrival times for more general spreading processes, including non-linear node-node interactions. Here, all paths between two given nodes contribute to the spreading time of signals, and the contribution of each path is further decomposed in terms of individual nodes' response time. These response time of each node is then linked to the specific dynamics at play, thus establishing a clear connection with global dynamical properties. The aforementioned methods provide a time-independent distance or quasi-distance function. 

Secondly, starting from the definition of network diffusion distance~\cite{coifman2005geometric,de2017diffusion}, a time-dependent geometrical framework for continuous diffusion is built and then extended to non-linear dynamics through the Jacobian distance \cite{barzon2024unraveling}. In the latter, a linearized dynamics of perturbations leads to a proper notion of time-dependent metric tensor, useful to probe mesoscopic patterns of signal spreading: the former can be obtained as a special case either by considering diffusive dynamics or by considering special nonlinear processes (e.g., Kuramoto or consensus dynamics) in specific regimes. Throughout this work, we highlight the salient mathematical aspects of these techniques in a pedagogical vein, and discuss results across related methods to facilitate due comparisons.

In the following, we will indicate with $N$ the number of nodes of a complex network $G(V,E)$, being $V$ the set of vertices or nodes, and $E$ the set of edges or connections among them. Furthermore, we indicate with $A_{ij}$ the corresponding adjacency matrix, indicating the presence or absence of a (possibly weighted) connection between nodes $i$ and $j$, and with $D_{ij}=k_{i}\delta_{ij}$ the node degree matrix, where $k_{i}=\sum_j A_{ij}$ denotes the degree of node $i$. Moreover, we will denote as $\mathbf{e}_i\in\mathbb{R}^{N}$ the $i$--th vector of the canonical base.

\section{Single scale dynamical distance measures}


A typical approach based on the choice of a specific time scale arises from linear cascade dynamics in the continuous time, as
\begin{eqnarray}
\dot{x}_{i}(t)=\sum_{j}A_{ij}x_{j}(t),
\end{eqnarray}
where $\mathbf{x}(t)$ denotes a state vector at time $t$. The solution yields a propagator $e^{\mathbf{A}t}$, and evaluating it at a specific time (e.g., $t=1$) recovers, for instance, the Estrada's communicability matrix~\cite{estrada2008communicability}, which can be used to define a distance~\cite{akbarzadeh2018communicability,boguna2021network}. 

Here, we will focus on more sophisticated approaches, involving nonlinear dynamics, and start by considering the effective distance induced by discrete time random walks originally introduced in \cite{brockmann2013hidden} and then expanded in \cite{iannelli2017effective,klamser2024inferring}. Next, we will turn our attention to more general definitions of dynamical distance measures, which can also account for the features of non-linear dynamics. In such cases, the interplay of network heterogeneity and non-linearity lead to a highly varied phenomenology, hindering general conclusions. This problem is first tackled in~\cite{hens2019spatiotemporal}, in which the dependence of signal propagation rate is shown to depend both on the network and dynamics. Building upon this framework, in~\cite{bontorin2023multi} the contribution of multiple paths is considered.

\subsection{Effective distance induced by a discrete time random walk}\label{sec:discrete_time_RW}

The main technical results of \cite{brockmann2013hidden} is the definition of an effective length for the path traveled by a random walker on a network. Let us denote with $p_i(t)$ the probability that the walker is at node $i$ at time $t$. We assume that the random walk is an homogeneous stochastic jump process, whose time evolution from time $t$ to time $t+1$ is described by the following equation:
\begin{equation}
    p_i(t+1)= \sum_j T_{ij} p_j(t),
\end{equation}
where $T_{ij}$ is a time-independent transition matrix, expressing the probability for the walker to jump from node $j$ to node $i$ in a discrete time step. If a stationary distribution $p_i(t=\infty)\equiv\pi_i$, such that
\begin{equation}
    \sum_j T_{ij}\pi_j= \pi_i \ ,
\end{equation}
also satisfies the detailed balance condition 
\begin{equation} \label{db_rw}
    T_{ij} \pi_j = T_{ji} \pi_i,
\end{equation}
then it is called an equilibrium distribution. From now on, we will work in the equilibrium case, fixing the time scale.

The purely topological characterization of a path length in the network, based on the number of edges $L$ that compose it, does not capture the dynamical features of the random walk. For an unweighted network, two distinct paths that start and end at the same nodes and have the same structural length $L$ are equivalent from a purely topological perspective. However, a random walker may prefer one path over the other, as encoded by the transition probabilities $T_{ij}$. To account for this dynamical selection, the following definition of effective length of the edge between two connected nodes $i$ and $j$ is proposed in \cite{brockmann2013hidden}:
\begin{equation} \label{eq:eff_dist}
    \ell_{ij}=1-\log(T_{ij}).
\end{equation}
Thus, the effective distance between nodes $i$ and $j$ increases as the transition probability between them decreases. It is important to note that $\ell_{ij}$ is not symmetric since, using the detailed balance condition (\ref{db_rw}), we have that $\ell_{ij}=\ell_{ji}-\log\left(\frac{\pi_i}{\pi_j}\right)$.
Then, given a path $\Gamma$ of topological length $L$ corresponding to an ordered set of connected nodes $\{\ n_0,...,n_L\}$, the effective length of $\Gamma$ can be defined as
\begin{equation}
    \lambda(\Gamma)\equiv \sum_{i=0}^{L-1}\ell_{n_{i+1}n_i}=L-\log[W(\Gamma)],
\end{equation}
where 
\begin{equation}
    W(\Gamma)\equiv\prod\limits_{i=0}^{L-1} T_{n_{i+1}n_i}
\end{equation}
is the probability for the random walker to traverse the path $\Gamma$. 

Now, we can exploit this definition of path effective length to project the network and its discrete diffusion dynamics in a latent space where the effective distance from a node $j$ to a node $i$ is obtained by minimizing the effective length $\lambda(\Gamma)$ on the set $\Omega_{ij}$ of all the paths going from $j$ to $i$:
\begin{equation} \label{eq:min_eff_dist}
    d_{ij}=\min\limits_{\Gamma \in \Omega_{ij}}\lambda(\Gamma).
\end{equation}

\begin{figure*}[htb]
        \centering
        \includegraphics[width=0.9\textwidth]{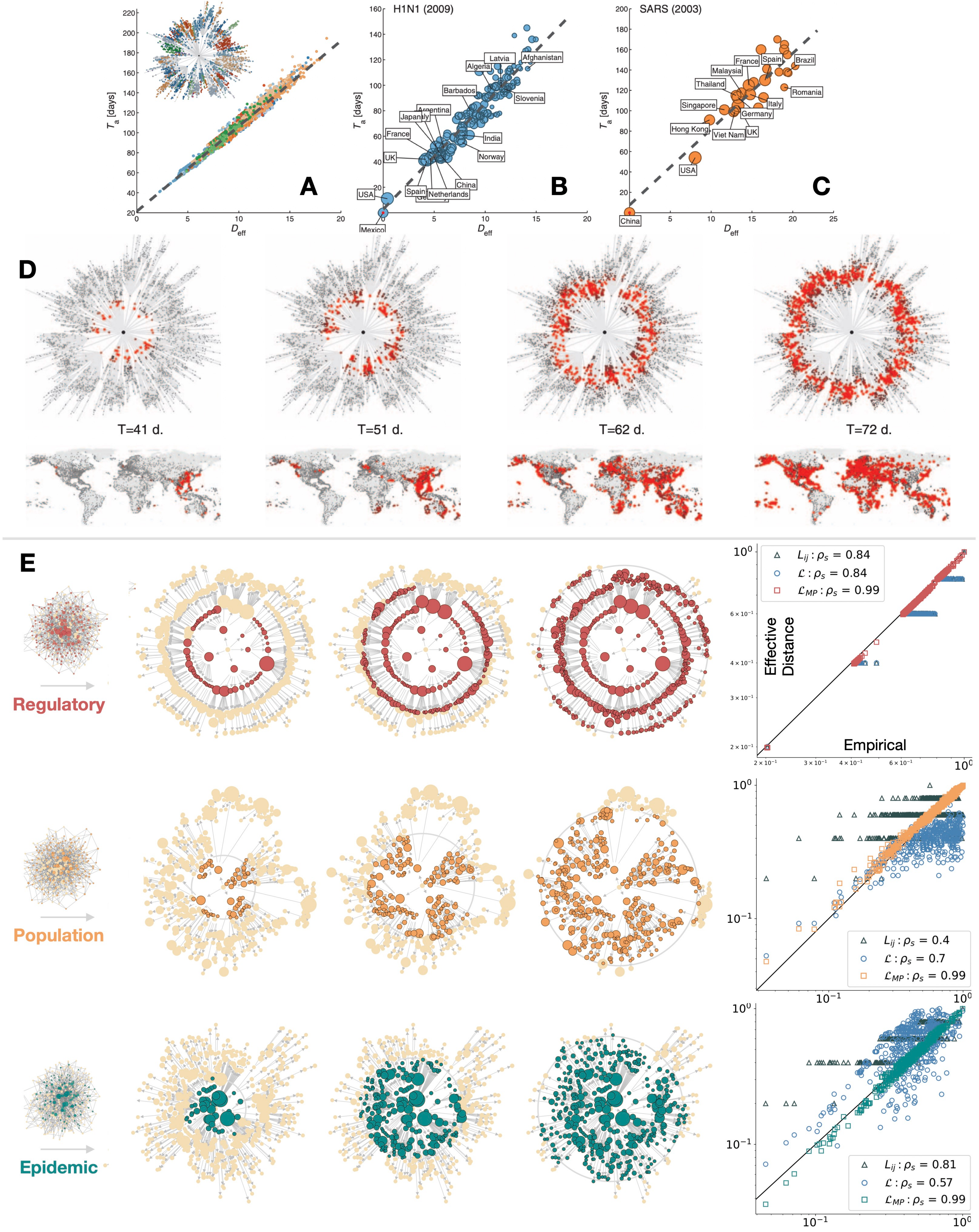}
\caption{\textbf{Effective distance for network dynamics.} In~\cite{brockmann2013hidden}, epidemic spreading on the World Airtransportation Network is shown to hide remarkably regular patterns which are unveiled by the effective distance (A). The introduction of an effective velocity for global spreading $V_{\textrm{eff}}$, dependent on a set of structural and epidemic parameters, allows the approximation of the arrival time at node $i$ as $\tau_{i\leftarrow j}= d_{ij} / V_{\textrm{eff}}$, where $j$ is the outbreak node. In (B-C) the comparison between effective distances and arrival times for H1N1 (2009) and SARS (2003) outbreaks \cite{brockmann2013hidden}. This approximation corresponds to a concentric front wave propagation with constant speed in the latent space (D). $\vert$ In~\cite{hens2019spatiotemporal,bontorin2023multi} propagation times for signals are computed by taking into account fundamentally different network dynamics (E), again finding regular patterns within dynamical classes. Correspondingly, a notion of distance emerges, where nodes are close if their respective propagation time is small (F). Panels A-D have been adapted from \cite{brockmann2013hidden}. Panel E has been adapted from \cite{bontorin2023multi}.}
    \label{fig:fig2}
\end{figure*}

This effective distance has proven to be a powerful tool to study meta-population models which feature a faster local process paired with a slower global diffusion, as shown in Fig.~\ref{fig:fig2} for a SIR epidemic.

This framework can be extended to build an effective distance that takes into account all the possible walks between two nodes, as it is done in \cite{iannelli2017effective}, and also the eventual dependence on some scale parameter $\delta$. This extension leads to more accurate estimates of arrival times. 

The first step towards this goal is the introduction of the scale parameter in the effective edge length (\ref{eq:eff_dist}):
\begin{equation}
    \ell_{ij}=\delta-\log(T_{ij}),
\end{equation}
from which we have the effective path length 
\begin{equation}
    \lambda(\Gamma,\delta)\equiv \sum_{i=0}^{L-1}\ell_{n_{i+1}n_i}=L\delta-\log[W(\Gamma)].
\end{equation}
This parametric definition of effective edge length has been recently used in \cite{klamser2024inferring} to fruitfully define a measure of import risk, linked to the probability for a given country to import a specific epidemic disease.

As second step, as shown in~\cite{gautreau2008global}, we write the effective length $\lambda(\Gamma+\Gamma')$ associated to the sum of two paths $\Gamma$ and $\Gamma'$ as
\begin{equation}
    e^{-\lambda(\Gamma+\Gamma')}=e^{-\lambda(\Gamma)}+e^{-\lambda(\Gamma')}.
\end{equation}
The aforementioned equation allows us to recover the overall effective distance associated to the set of paths $\{\Gamma_{ij}\}$ that go from a node $j$ to a node $i$ as 
\begin{equation}
    \lambda^{\{\Gamma_{ij}\}}_{ij}(\delta)=-\log \left[\sum_{\{\Gamma_{ij}\}}e^{-L(\Gamma_{ij})\delta}W(\Gamma_{ij}) \right],
\end{equation}
with $L(\Gamma_{ij})$ number of nodes in the path $\Gamma_{ij}$. If we define the probability to have a path of length $n$ as $F_{ij}(n)\equiv \sum_{\{\Gamma_{ij}\mid\ L(\Gamma_{ij})=n\}} W(\Gamma_{ij})$, then we can write
\begin{equation} \label{eq:multipath_eff_lenght}
    \lambda^{\{\Gamma_{ij}\}}_{ij}(\delta)= -\log \left[\sum_{n}e^{-n\delta} F_{ij}(n) \right],
\end{equation}
where the sum extends to the maximum possible structural length for a path in the network. We can further generalize \eqref{eq:multipath_eff_lenght} by considering all the possible walks between the starting node $j$ and the target node $i$, relaxing the constraint on the repetition of nodes, which is instead present in the definition of path. This gives us the possibility to rewrite \eqref{eq:multipath_eff_lenght} as
\begin{equation} \label{eq:gen_eff_dist}
    \Lambda_{ij}(\delta)= -\log \left[\sum_{n=0}^{\infty}e^{-n\delta} H_{ij}(n) \right],
\end{equation}
where $H_{ij}(n)$ is the probability to have a walk of length $n$ from $j$ to $i$\footnote{We have $H_{ij}(0)=0$ for $i\neq j$.}, which is related to the hitting time probability of the random walker. This identification allows us to deduce the following relation between the hitting time probability generating functional $C_{ij}(\delta)$ and the effective distance $\Lambda_{ij}(\delta)$:
\begin{equation}
    \Lambda_{ij}(\delta)=-C_{ij}(-\delta).
\end{equation}
Therefore, in a statistical physics perspective, we can interpret $\Lambda_{ij}(\delta)$ as a generalized free energy functional, obtaining a solid theoretical foundation for this definition of effective distance. The single path distance \eqref{eq:min_eff_dist} is recovered from \eqref{eq:gen_eff_dist} by discarding sub-leading contributions in the sum.

\subsection{Universal temporal distance for non-linear dynamics}

We consider a configuration model network endowed with the dynamics

\begin{equation}\label{eq:hens2019_dyn}
    \dot{x}_i = f(x_i) + \sum_{j=1}^N A_{ij}g(x_i,x_j) \ , 
\end{equation}

with the factorization assumption $g(z_1,z_2) \equiv g_{1}(z_1)g_{2}(z_2)$; this latter requirement is satisfied by many paradigmatic models of interacting systems, across several disciplines including epidemiology, ecology, population dynamics and neuroscience. We assume that a steady state is eventually reached, $\lim\limits_{t \to {\infty}}x_i = \bar{x}_i \ \forall i$, and perturb the steady state of node $j$ with a constant displacement: $\bar{x}_j \to \bar{x}_j + \Delta x_j$. Any given node $i$ will respond by relaxing to a new steady state, $x_i(t) = \bar{x}_i + \Delta x_i(t)$. The response time of node $i$ to the perturbation of $j$, $T(j\to i)$, is defined as the time necessary for $\Delta x_i(t)$ to reach a fraction $\eta$ of its asymptotic value:

\begin{equation}\label{eq:hens2019_propagation_time}
    \frac{\Delta x_i\left(t=T\left(j \to i\right)\right)}{\Delta x_i\left(t=\infty\right)} = \eta \ .
\end{equation}
In principle, this response time depends on $\eta$, but results hold qualitatively for $\eta \in (0,1)$. A sensible choice is $\eta = 1/2$, for which $T\left(j \to i\right)$ corresponds to the half-saturation time.

To derive analytical estimates of $T\left(j \to i\right)$ we first consider the relaxation time $\tau_i$ of node $i$ under a perturbation of its neighbor node $j$. $\tau_i$ is found in~\cite{hens2019spatiotemporal} to depend on the in-degree of node $i$, $k_i^{\textrm{in}} = \sum_{j}A_{ij}$, as 

\begin{equation}\label{eq:hen2019_relax_time}
   \tau_i \sim \left(k_i^{\textrm{in}}\right)^\theta \ , 
\end{equation}

where the scaling exponent $\theta$ depends only on the form of dynamics~\eqref{eq:hens2019_dyn} and can be determined analytically in a mean field approximation. Then, to go beyond first neighbors, and assuming that signals from $j$ to $i$ propagate mainly through the quickest (not shortest) path $\Gamma_{ij}$, we can estimate the propagation time as

\begin{equation}\label{eq:hens2019_propagation_time_shortest_path}
    T\left(j \to i\right) \sim \min_{\{\Gamma_{ij}\}} \sum_{\substack{l \in \Gamma_{ij} \\ l \neq j}} \tau_l .
\end{equation}

The overall propagation properties of a network-dynamics pair is described by the distribution of propagation times for all node pairs, $P(T)$, and its statistics, e.g. the mean propagation time $\left\langle T \right\rangle = \frac{1}{N(N-1)}\sum_{i,j}T\left(j \to i\right)\equiv \int T P(T) dT$. 

According to~\eqref{eq:hen2019_relax_time}, signal propagation under different dynamics can be classified according to the corresponding value of $\theta$:

\begin{itemize}
    \item[$\theta > 0$:] As $\partial_{k_i^{\textrm{in}}}\tau_i > 0$, hubs respond more slowly to perturbation, effectively slowing down signal propagation (\textbf{degree-limited propagation});
    \item[$\theta = 0$:] The response time of nodes is independent of their degrees, hence $T\left(j \to i\right) \sim \textrm{length}\left(\Gamma_{ij}\right)$ (\textbf{distance-limited propagation});
    \item[$\theta < 0$:] $\partial_{k_i^{\textrm{in}}}\tau_i < 0$, so that hubs accelerate propagation, which however is still limited by the overall length of $\Gamma_{ij}$ (\textbf{composite propagation}).
\end{itemize}

Strikingly, this classification correctly predicts that the scale-free property of networks has little effect on the propagation times for distance-limited propagation, which is bottle-necked by low-degrees nodes. Also, while the framework assumes a configuration model network, it correctly suggests that modularity leads to either extremely delayed propagation or immediate propagation in the cases of degree-limited and composite propagation, respectively. The profile of $P(T)$ is a more comprehensive (but still robust) pattern: $P_{\theta = 0}(T)$ shows regularly spaced peaks, $P_{\theta > 0}(T)$ is essentially unimodal, and $P_{\theta < 0}(T)$ shows overlapping peaks, each pertaining to paths of well-defined length.

\textbf{Multi-path corrections.} In general, the propagation time $T\left(j \to i\right)$ should take into account paths other than the quickest one, as explored e.g. in~\cite{iannelli2017effective} for epidemics spreading. In fact, the quickest path may be degenerate, i.e. it is not well-separated from all other possible paths in terms of their respective propagation times. Even in non-degenerate cases, the contribution of numerous secondary paths might considerably affect signal propagation. Correspondingly, the estimate in~\eqref{eq:hens2019_propagation_time_shortest_path} serves as a qualitative prediction of arrival times. In~\cite{bontorin2023multi}, these degenerate contributions are taken into account, leading to a multi-path estimate of propagation times, $T_{\textrm{MP}}\left(j \to i\right)$, which achieves excellent quantitative accuracy.

This extension considers the effect of a steady state constant perturbation of node $j$, $\bar{x}_j \to \bar{x}_j + \Delta x_j$, on the state of node $i$, $\bar{x}_i \to \bar{x}_i + \Delta x_i$. To account of the role of multiple routes, the authors of~\cite{bontorin2023multi} use the global correlation matrix as defined in~\cite{barzel2013universality}, namely

\begin{equation}\label{eq:bontorin2023_correlation_matrix}
    G_{ij} = \left|\frac{\Delta x_i / x_i}{\Delta x_j / x_j}\right|_{x=\bar x} \ .
\end{equation}

We note that $G_{ij}$ takes into account the full network structure. Further, $G$ is decomposed in terms of contributions on the ensemble of paths that connect source to target to track the amount of perturbation that is conveyed by each possible route:

\begin{equation}
    G_{ij} = \displaystyle\sum_{\{\Gamma_{ij}\}} G_{\Gamma_{ij}} \ .
    \label{eq:gij_assumption}
\end{equation}

The contribution of each path may be further broken down to local (i.e. nearest neighbors) correlations, encoded by the relative responses

\begin{equation}\label{eq:bontorin2023_local_correlation_matrix}
    R_{ij} = \left|\frac{\partial \log \left( x_i \right)}{\partial\log \left( x_j \right)}\right|_{x=\bar x} \ .
\end{equation}
 Therefore, for a given path of length $L$, $\Gamma_{ij} = \left(l_1,\dots,l_L\right)$ (where $l_1 = j$ and $l_L = i$), we may chain the corresponding sequence of local correlations, and write the path correlation as

\begin{equation}\label{eq:bontorin2023_path_correlation}
    G_{\Gamma_{ij}} = \prod_{i=1}^{L-1}R_{l_{i+1} \ l_i} \ .
\end{equation}

Equation \eqref{eq:bontorin2023_path_correlation} provides a way to compute single-path contributions. In turn, we can define the propagation weight of each path between $j$ and $i$ as $W_{\Gamma_{ij}} = G_{\Gamma_{ij}}/G_{ij}$.


Given the single-path propagation time $\tau_{\Gamma_{ij}} = \sum_{l \in \Gamma_{ij}}\tau_l$ (with $\tau_l$ as in~\eqref{eq:hen2019_relax_time}), we can write the multi-path propagation time as

\begin{equation}\label{eq:bontorin2023_multipath_time}
    T_{MP}\left(j \to i\right) = \sum_{\{\Gamma_{ij}\}} W_{\Gamma_{ij}} \tau_{\Gamma_{ij}} \ .
\end{equation}

We remark that both~\eqref{eq:hens2019_propagation_time_shortest_path} and~\eqref{eq:bontorin2023_multipath_time} serve as effective pseudo-distances: node pairs with short propagation times are naturally considered as closer than pairs with long propagation times.


\section{Multi-resolution dynamical distance measures}

The distances defined in the previous section are limited to the study of the network at a fixed time scale, which can be interpreted as probing network functionality at a given temporal resolution. Note that allowing a dynamical process to unfold on the top of a complex network allows one to inspect distinct topological scales by tuning the time as a resolution parameter. At low time scales, only local structural features are responsible for functionalities, while at high time scales the integration between distinct structural features is captured and can be used to inspect, for instance, the underlying mesoscopic organization.

\begin{figure*}[htb]
        \centering
        \includegraphics[width=1.0\textwidth]{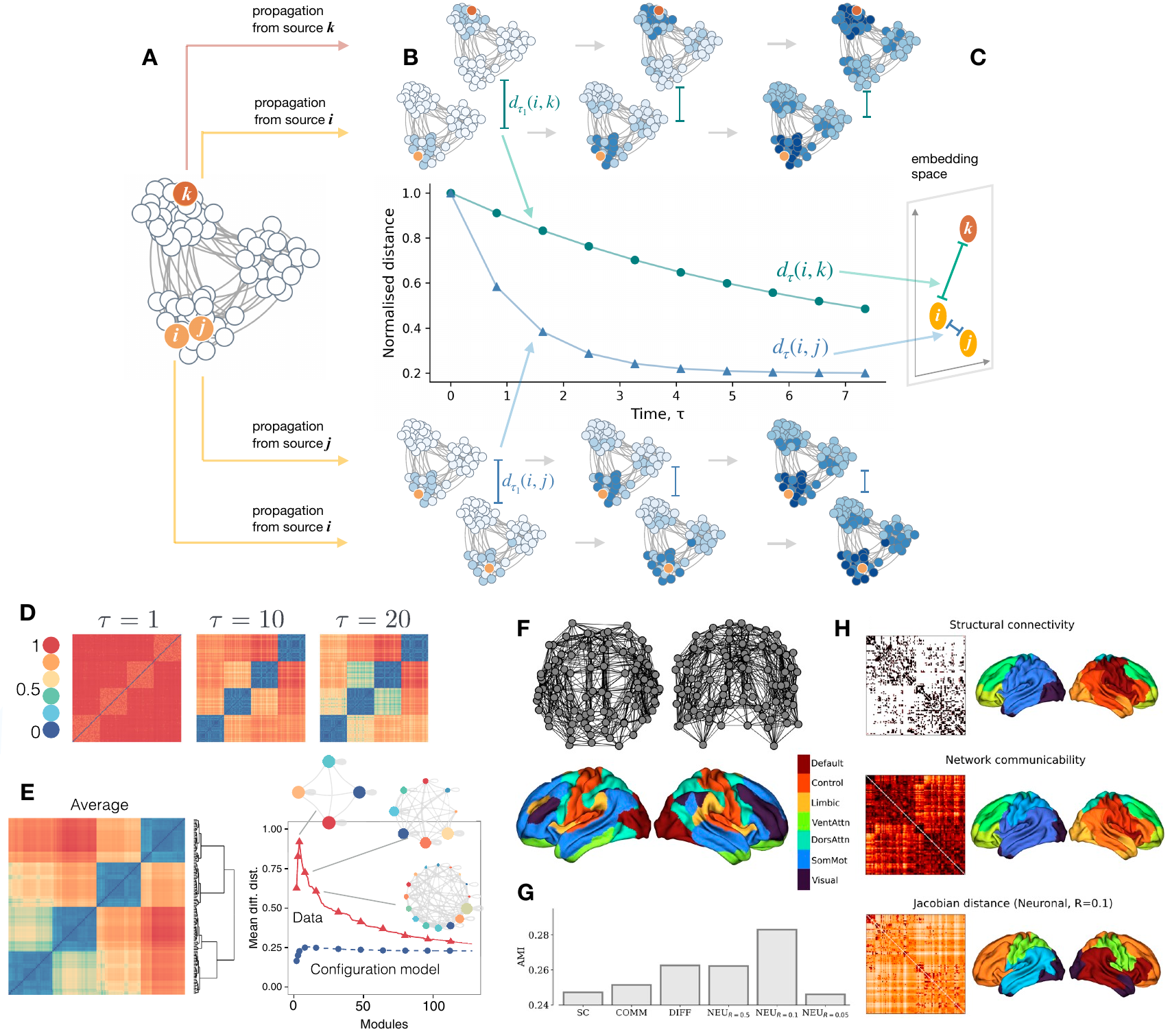}
\caption{\textbf{Multi-scale distances}. (A) Network propagation states emerge from different source nodes. Pairs of nodes are assigned a time-dependent distance (B) according to the difference between their propagation states at a time $\tau$, thus unveiling different mesoscopic structures, with distances between nodes in different communities shrinking over longer time scales with respect to nodes in the same community; thus, an time-dependent embedding space emerges (C).
In~\cite{de2017diffusion} the distance is computed according to the network states defined by the probability of being reached by a random walker starting at one of the source nodes $i$ or $j$. The set of pairwise diffusion distances results in a distance matrix across temporal scales (D) which can be then averaged (E) and compared with a configuration model to reveal a meaningful mesoscale structure in the diffusion process. (F) In~\cite{barzon2024unraveling}, empirical neural networks and functional brain mappings are analyzed through the Jacobian distance induced by a model of neural dynamics. (G) Adjusted mutual information (AMI) is optimized by the modular structure predicted by the Jacobian distance, which predicts more integrated modules than structural connectivity and network communicability (H). Panels D-E have been adapted from \cite{de2017diffusion}. Panels F-H have been adapted from \cite{barzon2024unraveling}.}
    \label{fig:fig3}
\end{figure*}

In this section, we consider the framework introduced in~\cite{de2017diffusion} for continuous time random walks, inspired by diffusion maps~\cite{coifman2005geometric}, and then extended in \cite{barzon2024unraveling} to small perturbations around a fixed point of a non-linear dynamics, which enables the definition of a multi-resolution distance measure for the network dynamics (see Fig.~\ref{fig:fig3}).

\subsection{Continuous-time diffusion}\label{sec:cont-time_diffusion}
The idea of associating an effective distance with a discrete random walk discussed in Section~\ref{sec:discrete_time_RW} can be extended to continuous random walks on networks, as it has been done~\cite{de2017diffusion} exploiting the diffusion distance described in~\cite{belkin2001laplacian}.

In the case of a continuous time random walk, the probability $p_i(t)$ for the walker to be in node $i$ at time $t$ evolves according to the master equation 
\begin{equation}
    \dot{p}_i(t)=-\sum_j \Tilde{\mathbf{L}}_{ij} p_j(t),
\end{equation}
where $\Tilde{\mathbf{L}}_{ij}$ is the normalized Laplacian of the network, defined as $\tilde{\mathbf{L}} = \mathbf{I} - \mathbf{D}^{-1}\mathbf{A}$, with $D= \text{diag}\left(k_1,\dots,k_N\right)$. The solution of this master equation is 
\begin{equation}
    \boldsymbol{p}(t)= e^{-t \Tilde{\mathbf{L}}} \boldsymbol{p}(0),
\end{equation}
in which $\boldsymbol{p}(t)$ is the vector of probabilities for the walker to be in a given node at time $t$. In the following,
$\boldsymbol{p}(t|i)$ denotes the probability vector obtained by imposing that the walker starts in node $i$, $\boldsymbol{p}(0)=\boldsymbol{e}_i$. Then, it is possible to define the diffusion distance $d_t(i,j)$ between two nodes $i$ and $j$ as the Euclidean distance between the the corresponding vectors $\boldsymbol{p}(t|i)$ and $\boldsymbol{p}(t|j)$:
\begin{equation}\label{diffusion_dist}
    d^2_{t}(i,j)=\left\lVert\boldsymbol{p}(t|i)-\boldsymbol{p}(t|j)\right\rVert^2= \left\lVert e^{-t \tilde{\mathbf{L}}} (\boldsymbol{e}_i-\boldsymbol{e}_j)\right\rVert^2.
\end{equation}
As stressed by the last equality in (\ref{diffusion_dist}), the diffusion distance can be seen as inducing a new non-Euclidean metric on the vector space of initial conditions, given by
\begin{equation} \label{diffusion metric}
    g_{kl}=[(e^{-t \Tilde{\mathbf{L}}})^T(e^{-t \Tilde{\mathbf{L}}})]_{kl}.
\end{equation}
Indeed, (\ref{diffusion metric}) satisfies all the requirements to be a metric tensor. 

A diffusion distance, as defined in \eqref{diffusion_dist}, measures how easily two nodes exchange information: the greater the number of paths connecting two nodes, the shorter will be their diffusion distance (Fig.~\ref{fig:fig3}A--C). For this reason, it can be used to probe the functional structure of a network on which a diffusion-like dynamic is defined, as done in \cite{de2017diffusion} for synchronization near a metastable state and consensus dynamics. In particular, this framework can be used both to identify functional clusters at a fixed timescale $\tau$ and to analyze the evolution and persistence of the mesoscale.
Indeed, from \eqref{diffusion_dist}, we can define the matrix of distances between the nodes $\Delta_{ij}$. The entries of this matrix tend to decrease with time, thus we introduce its normalized version
\begin{equation}\label{eq:MDD2017_temporal_distance_1time}
    \Tilde{\Delta}_{ij}(\tau)=\frac{\Delta_{ij}(\tau)}{\max_{ij}[\Delta_{ij}(\tau)]}
\end{equation}
to allow the comparison of the cluster structure between different time-scales. This matrix encodes the persistence of mesoscale as the persistence of geometrical distance between nodes of different clusters. This ensures that the time-averaged diffusion distance 
\begin{equation}\label{eq:MDD2017_temporal_distance_RW}
    \overline{\Delta}=\frac{1}{\tau_{max}} \sum_{\tau}^{\tau_{max}}\Tilde{\Delta}(\tau)
\end{equation}
serves as a measure of the network functional structure at a given cutoff time-scale $\tau_{max}$. Specifically, in a network with $N$ nodes, if $\tau_{max} \ll N$, then the diffusion is limited and the diffusion distance reflects only the local structure, while, if $\tau_{max}\approx N$, the mesoscale structure emerges in $\overline{\Delta}$, as the diffusion process has enough time to explore the entire network. Finally, in the case where $\tau_{max}\gg N$, the matrix retains only macroscale information. This is due to the fact that the different time-scales over which the network is explored by diffusion are inversely proportional to the eigenvalues of $\Tilde{\mathbf{L}}$, and the eigenvalues associated with the mesoscale dynamics approximately scale as $\lambda \approx \frac{1}{N}$~\cite{mohar1991laplacian}.


It is worth noticing that some structural indicators which have traditionally been the basis for models and analysis of complex networks, such as centrality descriptors and rich club organization, can be redefined or reinterpreted in terms of information carried by dynamics~\cite{bertagnolli2022functional}. Moreover, diffusion geometry can be generalized to be used with distinct flavors of random walk processes and can be suitably generalized to analyze more complex topologies, such as multiplex networks~\cite{bertagnolli2021diffusion}.

\subsection{Jacobian metric induced by linearized dynamics}\label{sec:jacobian}


The results of the previous section are extended in~\cite{barzon2024unraveling} to non-linear dynamics by tracking the propagation of small perturbations. Specifically, let us consider the dynamics $\dot{x}_i = f_{i}\left(x_1,\dots,x_N\right)$; the network structure is implicitly contained in the definition of $f$. Assuming the existence of a stable stationary state $\mathbf{\bar{x}} = \left(\bar{x}_1,\dots,\bar{x}_N\right)$, the linearized dynamics around this steady state read $\mathbf{\delta \dot{x}} = \mathbf{M} \mathbf{\delta x}$, where $\mathbf{M} = \left[\left.\frac{\partial f_i}{\partial x_j}\right|_{\mathbf{\bar{x}}}\right]_{ij}$, $\mathbf{\delta x} = \mathbf{x}(t) -\mathbf{\bar{x}}$ and with solution 
\begin{equation}\label{eq:ev_pert}
    \mathbf{\delta x}(t) = e^{t\mathbf{M}}\mathbf{\delta x}(0).
\end{equation}
 The Jacobian metric is built through the time evolution of single-site perturbations. More specifically, we define the distance between nodes $i$ and $j$ at time $\tau$ as

\begin{equation}\label{eq:barzon2024_jacobian_distance}
d_{\tau}\left(i,j\right) =  \left\lVert e^{\tau \mathbf{M}}\left[ \Delta x_i \mathbf{e}_i - \Delta x_j \mathbf{e}_j \right] \right\rVert  \ .
\end{equation}

Here, $\Delta x_i$ represents the scale of the initial perturbation of node $i$, which then evolves under the linearized dynamics. This definition applies both to diffusive systems and systems with non-linear node-node interactions, thus unifying previous approaches and generalizing the distance defined in Eq.~\eqref{diffusion_dist}. We remark, however, that Eq.~\eqref{diffusion_dist} entails no approximation, as it is built upon the master equation which is exactly linear. Nevertheless, at any time $\tau$, $d_{\tau}\left(i,j\right)$ defined in Eq.~\eqref{eq:barzon2024_jacobian_distance} is also a proper metric, being manifestly symmetric, non-negative and satisfying the triangular inequality. 

Similarly to the approach illustrated in Fig.~\ref{fig:fig3}A--C for diffusion geometry, the most persistent mesoscale patterns can be captured by averaging the network state over a sequence of times, $\left(\tau_1,\dots,\tau_\textrm{max}\right)$, with $0 < \tau_1 <\tau_\textrm{max} \approx N$, as described in Section~\ref{sec:cont-time_diffusion}, Eq.~\eqref{eq:MDD2017_temporal_distance_RW} (Fig.~\ref{fig:fig3}D--F).



\section{Outlooks} \label{sec:outlooks}

\begin{figure*}[ht]
        \centering
        \includegraphics[width=1.0\textwidth]{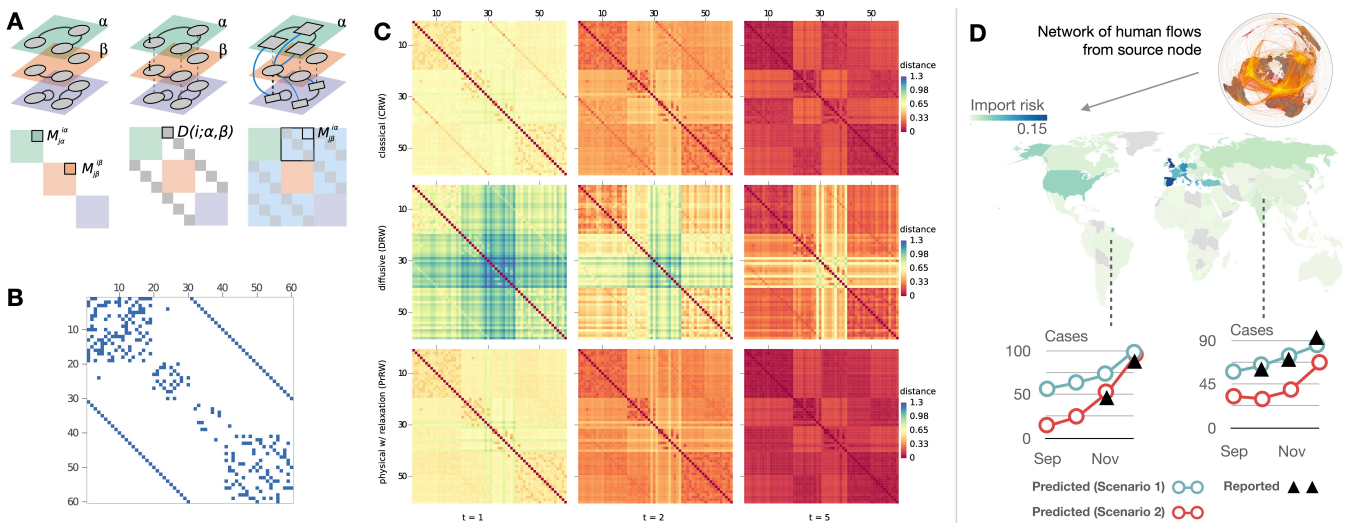}
\caption{\textbf{Applications of latent network geometry}. In~\cite{bertagnolli2021diffusion}, diffusion metrics of multi-layer networks is explored through the lenses of classical, diffusive and physical with relaxation random walks, building phenomenological models for the emerging geometry of a broad class of complex phenomena. The supra-adjacency matrix (A-B) defines the multiplex structure of a network, whereas the supra-distance matrices (C), taken at different times, reveal network substructures at different scales and of different types as explored by different families of random walkers. $\vert$ (D) In \cite{klamser2023enhancing} the latent geometry induced by the spreading a pathogen on the World Airtransportation Network from an outbreak airport is exploited to estimate the import risk \cite{klamser2024inferring} of an infected individual. This effective distance on a human mobility network enters as a part of a pipeline to infer the pandemic potential of a Variant of Concern. Panels A-C have been adapted from \cite{bertagnolli2021diffusion}. Panel D has been adapted from \cite{klamser2023enhancing}.}
    \label{fig:fig4}
\end{figure*}

The study of latent geometry emerging from network-driven processes found several recent applications to network modeling and analysis (see also Fig.~\ref{fig:fig4}). As this framework has developed only in relatively recent times, several questions -- which lie at the intersection of network science and other  branches of mathematics -- remain unexplored. Here, we outline some such intersectional lines of research which are of intrinsic mathematical and practical interest, potentially leading to novel insights in real world complex systems.

\vspace{0.5cm}

\emph{Dynamical steady states and rare-region effects.} The methods shown in  Section~\ref{sec:jacobian} assume the existence of a stable fixed point. However, the development of a Jacobian-like metric for system which do not necessarily reach constant steady states is of considerable interest. For example, oscillations in dynamical steady states could be approximated by self-consistent noise sources~\cite{kati2024self}, leading to the definition of stochastically fluctuating distances between nodes, and characterizing signal propagation through non-stationary systems. Moreover, the existence of chimera states in oscillating systems~\cite{davidsen2024introduction}, characterized by partial synchronization, could well lead to the emergence of exotic geometries, and anomalous signal propagation. These developments will enhance our understanding of phenomena such as neural signals' transmission~\cite{wang2021effect}.

Similarly, it will be of great interest to investigate the geometry induced by Griffiths phases on heterogeneous networks~\cite{munoz2010griffiths}. In Griffiths phases, due to rare regions of coherent disorder, one observes anomalous relaxation times. Correspondingly, signal propagation will most likely exhibit equally anomalous behavior. Approaching this scenario with the techniques of~\cite{hens2019spatiotemporal,bontorin2023multi} has the potential to unlock novel insights into long-lived transients in complex systems, with particular relevance to ecological forecasts~\cite{hastings2018transient}.

\vspace{0.5cm}

\emph{Parallels with fluid dynamics.} The flow of a fluid described by a first-order differential equation of the type $\dot{\mathbf{x}}(t)=\mathbf{f}(\mathbf{x})$ can be characterized by the presence of Lagrangian Coherent Structures \cite{lekien2007lagrangian,haller2015lagrangian}. These are dynamically emerging surfaces of trajectories that temporarily determine the skeleton of the flow, effectively dividing it into distinct dynamical regions. One of the most useful tools to study Lagrangian Coherent Structures is the Cauchy-Green tensor, which measures the deformation induced by the flow with respect to the initial conditions, and can be used to find Lagrangian Coherent Structures thorough variational methods \cite{haller2011variational}. A similar treatment can be extended to the flow of complex networks in their state space, in order to identify the geography of dynamical regions corresponding to the network at a given time and eventual barriers between them. A first step in this direction is the definition of the Jacobian metric tensor of \cite{barzon2024unraveling}, which can be interpreted as the Cauchy-Green tensor of the linearized flow.

\vspace{0.5cm}

\emph{Integration between information and network geometry.} The application of Information geometry methods \cite{amari2016information} has enriched many fields, among which Thermodynamics, Control Theory, and Statistical and Quantum Mechanics \cite{janke2002information,kolodrubetz2017geometry,erdmenger2020information,floerchinger2023information,kim2021information,ito2018stochastic}. In fact, the possibility to associate a differentiable manifold to a probability distribution has paved the way for the exploitation of a wide array of differential geometry techniques, in particular to characterize the phase structure of various physical systems \cite{janke2004information,carollo2020geometry,banchi2014quantum}. This has been done also for a complex network endowed with a Kuramoto dynamics \cite{alexandrov2023information}, linking the critical point of the synchronization phase transition with the singularity of the Fisher information metric tensor associated to the system. Despite these advancements, a comprehensive information-geometrical exploration of phase transitions in complex networks is still lacking and would greatly enhance their understanding.

\vspace{0.5cm}

\emph{Dynamical Mean-Field Theory of network dynamics and vestigial geometry.} Recently, a description of dynamics in complex networks in terms of state path probability provided by Dynamical Mean-Field Theory \cite{metz2025dynamical} has been achieved, although with some limitations. As the theory is in rapid development, a cross-contamination of the two frameworks seems inevitable. On the one hand, dynamical mean-field theories can furnish a description of background noise by determining the fluctuations of a single node in a stationary (i.e. not externally perturbed) regime. On the other hand, given a finite-size networked dynamical system, the latent geometry approach might shed some light on the self-consistent bath coupled to the effective degree of freedom, which in the general case is a famously complicated object to deal with.

\vspace{0.5cm}

Overall, the possibility to study network functionality in terms of microscopic and mesoscale organization based on generative models, is tantalizing. Our proposed taxonomy of latent geometries emerging from network‐driven processes, contrasting fixed‐time and multi‐scale reconstructions, highlight their mathematical foundations and applications in revealing functional organization across complex systems and can be used to better characterize future developments in this field.

\section{Acknowledgments}

AFB and MDD acknowledge MUR funding within the FIS (DD n. 1219 31-07-2023) Project no. FIS00000158 (CUP C53C23000660001). DZ acknowledges the Human Frontier Science Program Organization (HFSP Ref. RGY0064/2022). SB acknowledges the support of the PNRR ICSC National Research Centre for High Performance Computing, Big Data and Quantum Computing (CN00000013), under the NRRP MUR program funded by the NextGenerationEU.

\bibliographystyle{naturemag}
\bibliography{biblio}{}

\begin{thebibliography}{10}
\expandafter\ifx\csname url\endcsname\relax
  \def\url#1{\texttt{#1}}\fi
\expandafter\ifx\csname urlprefix\endcsname\relax\def\urlprefix{URL }\fi
\providecommand{\bibinfo}[2]{#2}
\providecommand{\eprint}[2][]{\url{#2}}

\bibitem{newman2003structure}
\bibinfo{author}{Newman, M.~E.}
\newblock \bibinfo{title}{The structure and function of complex networks}.
\newblock \emph{\bibinfo{journal}{SIAM review}} \textbf{\bibinfo{volume}{45}}, \bibinfo{pages}{167--256} (\bibinfo{year}{2003}).

\bibitem{boccaletti2006complex}
\bibinfo{author}{Boccaletti, S.}, \bibinfo{author}{Latora, V.}, \bibinfo{author}{Moreno, Y.}, \bibinfo{author}{Chavez, M.} \& \bibinfo{author}{Hwang, D.-U.}
\newblock \bibinfo{title}{Complex networks: Structure and dynamics}.
\newblock \emph{\bibinfo{journal}{Physics reports}} \textbf{\bibinfo{volume}{424}}, \bibinfo{pages}{175--308} (\bibinfo{year}{2006}).

\bibitem{dorogovtsev2008critical}
\bibinfo{author}{Dorogovtsev, S.~N.}, \bibinfo{author}{Goltsev, A.~V.} \& \bibinfo{author}{Mendes, J.~F.}
\newblock \bibinfo{title}{Critical phenomena in complex networks}.
\newblock \emph{\bibinfo{journal}{Reviews of Modern Physics}} \textbf{\bibinfo{volume}{80}}, \bibinfo{pages}{1275--1335} (\bibinfo{year}{2008}).

\bibitem{gao2012networks}
\bibinfo{author}{Gao, J.}, \bibinfo{author}{Buldyrev, S.~V.}, \bibinfo{author}{Stanley, H.~E.} \& \bibinfo{author}{Havlin, S.}
\newblock \bibinfo{title}{Networks formed from interdependent networks}.
\newblock \emph{\bibinfo{journal}{Nature physics}} \textbf{\bibinfo{volume}{8}}, \bibinfo{pages}{40--48} (\bibinfo{year}{2012}).

\bibitem{holme2012temporal}
\bibinfo{author}{Holme, P.} \& \bibinfo{author}{Saram{\"a}ki, J.}
\newblock \bibinfo{title}{Temporal networks}.
\newblock \emph{\bibinfo{journal}{Physics reports}} \textbf{\bibinfo{volume}{519}}, \bibinfo{pages}{97--125} (\bibinfo{year}{2012}).

\bibitem{liu2016control}
\bibinfo{author}{Liu, Y.-Y.} \& \bibinfo{author}{Barab{\'a}si, A.-L.}
\newblock \bibinfo{title}{Control principles of complex systems}.
\newblock \emph{\bibinfo{journal}{Reviews of Modern Physics}} \textbf{\bibinfo{volume}{88}}, \bibinfo{pages}{035006} (\bibinfo{year}{2016}).

\bibitem{battiston2020networks}
\bibinfo{author}{Battiston, F.} \emph{et~al.}
\newblock \bibinfo{title}{Networks beyond pairwise interactions: Structure and dynamics}.
\newblock \emph{\bibinfo{journal}{Physics reports}} \textbf{\bibinfo{volume}{874}}, \bibinfo{pages}{1--92} (\bibinfo{year}{2020}).

\bibitem{rosas2022disentangling}
\bibinfo{author}{Rosas, F.~E.} \emph{et~al.}
\newblock \bibinfo{title}{Disentangling high-order mechanisms and high-order behaviours in complex systems}.
\newblock \emph{\bibinfo{journal}{Nature Physics}} \textbf{\bibinfo{volume}{18}}, \bibinfo{pages}{476--477} (\bibinfo{year}{2022}).

\bibitem{de2023more}
\bibinfo{author}{De~Domenico, M.}
\newblock \bibinfo{title}{More is different in real-world multilayer networks}.
\newblock \emph{\bibinfo{journal}{Nature Physics}} \textbf{\bibinfo{volume}{19}}, \bibinfo{pages}{1247--1262} (\bibinfo{year}{2023}).

\bibitem{gross2008adaptive}
\bibinfo{author}{Gross, T.} \& \bibinfo{author}{Blasius, B.}
\newblock \bibinfo{title}{Adaptive coevolutionary networks: a review}.
\newblock \emph{\bibinfo{journal}{Journal of the Royal Society Interface}} \textbf{\bibinfo{volume}{5}}, \bibinfo{pages}{259--271} (\bibinfo{year}{2008}).

\bibitem{perc2013evolutionary}
\bibinfo{author}{Perc, M.}, \bibinfo{author}{G{\'o}mez-Gardenes, J.}, \bibinfo{author}{Szolnoki, A.}, \bibinfo{author}{Flor{\'\i}a, L.~M.} \& \bibinfo{author}{Moreno, Y.}
\newblock \bibinfo{title}{Evolutionary dynamics of group interactions on structured populations: a review}.
\newblock \emph{\bibinfo{journal}{Journal of the royal society interface}} \textbf{\bibinfo{volume}{10}}, \bibinfo{pages}{20120997} (\bibinfo{year}{2013}).

\bibitem{berner2023adaptive}
\bibinfo{author}{Berner, R.}, \bibinfo{author}{Gross, T.}, \bibinfo{author}{Kuehn, C.}, \bibinfo{author}{Kurths, J.} \& \bibinfo{author}{Yanchuk, S.}
\newblock \bibinfo{title}{Adaptive dynamical networks}.
\newblock \emph{\bibinfo{journal}{Physics Reports}} \textbf{\bibinfo{volume}{1031}}, \bibinfo{pages}{1--59} (\bibinfo{year}{2023}).

\bibitem{sole2024fundamental}
\bibinfo{author}{Sol{\'e}, R.} \emph{et~al.}
\newblock \bibinfo{title}{Fundamental constraints to the logic of living systems}.
\newblock \emph{\bibinfo{journal}{Interface Focus}} \textbf{\bibinfo{volume}{14}}, \bibinfo{pages}{20240010} (\bibinfo{year}{2024}).

\bibitem{de2025architecture}
\bibinfo{author}{De~Domenico, M.}
\newblock \bibinfo{title}{The architecture of living systems}.
\newblock \emph{\bibinfo{journal}{Under Review}}  (\bibinfo{year}{2025}).

\bibitem{barzel2013universality}
\bibinfo{author}{Barzel, B.} \& \bibinfo{author}{Barab{\'a}si, A.-L.}
\newblock \bibinfo{title}{Universality in network dynamics}.
\newblock \emph{\bibinfo{journal}{Nature physics}} \textbf{\bibinfo{volume}{9}}, \bibinfo{pages}{673--681} (\bibinfo{year}{2013}).

\bibitem{hens2019spatiotemporal}
\bibinfo{author}{Hens, C.}, \bibinfo{author}{Harush, U.}, \bibinfo{author}{Haber, S.}, \bibinfo{author}{Cohen, R.} \& \bibinfo{author}{Barzel, B.}
\newblock \bibinfo{title}{Spatiotemporal signal propagation in complex networks}.
\newblock \emph{\bibinfo{journal}{Nature Physics}} \textbf{\bibinfo{volume}{15}}, \bibinfo{pages}{403--412} (\bibinfo{year}{2019}).

\bibitem{de2016spectral}
\bibinfo{author}{De~Domenico, M.} \& \bibinfo{author}{Biamonte, J.}
\newblock \bibinfo{title}{Spectral entropies as information-theoretic tools for complex network comparison}.
\newblock \emph{\bibinfo{journal}{Physical Review X}} \textbf{\bibinfo{volume}{6}}, \bibinfo{pages}{041062} (\bibinfo{year}{2016}).

\bibitem{edler2017mapping}
\bibinfo{author}{Edler, D.}, \bibinfo{author}{Bohlin, L.} \& \bibinfo{author}{Rosvall, M.}
\newblock \bibinfo{title}{Mapping higher-order network flows in memory and multilayer networks with infomap}.
\newblock \emph{\bibinfo{journal}{Algorithms}} \textbf{\bibinfo{volume}{10}}, \bibinfo{pages}{112} (\bibinfo{year}{2017}).

\bibitem{ghavasieh2020statistical}
\bibinfo{author}{Ghavasieh, A.}, \bibinfo{author}{Nicolini, C.} \& \bibinfo{author}{De~Domenico, M.}
\newblock \bibinfo{title}{Statistical physics of complex information dynamics}.
\newblock \emph{\bibinfo{journal}{Physical Review E}} \textbf{\bibinfo{volume}{102}}, \bibinfo{pages}{052304} (\bibinfo{year}{2020}).

\bibitem{ghavasieh2023generalized}
\bibinfo{author}{Ghavasieh, A.} \& \bibinfo{author}{De~Domenico, M.}
\newblock \bibinfo{title}{Generalized network density matrices for analysis of multiscale functional diversity}.
\newblock \emph{\bibinfo{journal}{Physical Review E}} \textbf{\bibinfo{volume}{107}}, \bibinfo{pages}{044304} (\bibinfo{year}{2023}).

\bibitem{villegas2023laplacian}
\bibinfo{author}{Villegas, P.}, \bibinfo{author}{Gili, T.}, \bibinfo{author}{Caldarelli, G.} \& \bibinfo{author}{Gabrielli, A.}
\newblock \bibinfo{title}{Laplacian renormalization group for heterogeneous networks}.
\newblock \emph{\bibinfo{journal}{Nature Physics}} \textbf{\bibinfo{volume}{19}}, \bibinfo{pages}{445--450} (\bibinfo{year}{2023}).

\bibitem{peixoto2025network}
\bibinfo{author}{Peixoto, T.~P.}
\newblock \bibinfo{title}{Network reconstruction via the minimum description length principle}.
\newblock \emph{\bibinfo{journal}{Physical Review X}} \textbf{\bibinfo{volume}{15}}, \bibinfo{pages}{011065} (\bibinfo{year}{2025}).

\bibitem{zhang2025coarse}
\bibinfo{author}{Zhang, Z.}, \bibinfo{author}{Ghavasieh, A.}, \bibinfo{author}{Zhang, J.} \& \bibinfo{author}{De~Domenico, M.}
\newblock \bibinfo{title}{Coarse-graining network flow through statistical physics and machine learning}.
\newblock \emph{\bibinfo{journal}{Nature Communications}} \textbf{\bibinfo{volume}{16}}, \bibinfo{pages}{1605} (\bibinfo{year}{2025}).

\bibitem{rosvall2008maps}
\bibinfo{author}{Rosvall, M.} \& \bibinfo{author}{Bergstrom, C.~T.}
\newblock \bibinfo{title}{Maps of random walks on complex networks reveal community structure}.
\newblock \emph{\bibinfo{journal}{Proceedings of the national academy of sciences}} \textbf{\bibinfo{volume}{105}}, \bibinfo{pages}{1118--1123} (\bibinfo{year}{2008}).

\bibitem{rosvall2009map}
\bibinfo{author}{Rosvall, M.}, \bibinfo{author}{Axelsson, D.} \& \bibinfo{author}{Bergstrom, C.~T.}
\newblock \bibinfo{title}{The map equation}.
\newblock \emph{\bibinfo{journal}{The European Physical Journal Special Topics}} \textbf{\bibinfo{volume}{178}}, \bibinfo{pages}{13--23} (\bibinfo{year}{2009}).

\bibitem{viamontes2011compression}
\bibinfo{author}{Viamontes~Esquivel, A.} \& \bibinfo{author}{Rosvall, M.}
\newblock \bibinfo{title}{Compression of flow can reveal overlapping-module organization in networks}.
\newblock \emph{\bibinfo{journal}{Physical Review X}} \textbf{\bibinfo{volume}{1}}, \bibinfo{pages}{021025} (\bibinfo{year}{2011}).

\bibitem{lambiotte2015random}
\bibinfo{author}{Lambiotte, R.}, \bibinfo{author}{Delvenne, J.-C.} \& \bibinfo{author}{Barahona, M.}
\newblock \bibinfo{title}{Random walks, markov processes and the multiscale modular organization of complex networks}.
\newblock \emph{\bibinfo{journal}{IEEE Transactions on Network Science and Engineering}} \textbf{\bibinfo{volume}{1}}, \bibinfo{pages}{76--90} (\bibinfo{year}{2015}).

\bibitem{nicolini2017community}
\bibinfo{author}{Nicolini, C.}, \bibinfo{author}{Bordier, C.} \& \bibinfo{author}{Bifone, A.}
\newblock \bibinfo{title}{Community detection in weighted brain connectivity networks beyond the resolution limit}.
\newblock \emph{\bibinfo{journal}{Neuroimage}} \textbf{\bibinfo{volume}{146}}, \bibinfo{pages}{28--39} (\bibinfo{year}{2017}).

\bibitem{ghavasieh2021unraveling}
\bibinfo{author}{Ghavasieh, A.}, \bibinfo{author}{Stella, M.}, \bibinfo{author}{Biamonte, J.} \& \bibinfo{author}{De~Domenico, M.}
\newblock \bibinfo{title}{Unraveling the effects of multiscale network entanglement on empirical systems}.
\newblock \emph{\bibinfo{journal}{Communications Physics}} \textbf{\bibinfo{volume}{4}}, \bibinfo{pages}{129} (\bibinfo{year}{2021}).

\bibitem{villegas2022laplacian}
\bibinfo{author}{Villegas, P.}, \bibinfo{author}{Gabrielli, A.}, \bibinfo{author}{Santucci, F.}, \bibinfo{author}{Caldarelli, G.} \& \bibinfo{author}{Gili, T.}
\newblock \bibinfo{title}{Laplacian paths in complex networks: Information core emerges from entropic transitions}.
\newblock \emph{\bibinfo{journal}{Physical Review Research}} \textbf{\bibinfo{volume}{4}}, \bibinfo{pages}{033196} (\bibinfo{year}{2022}).

\bibitem{bovet2022flow}
\bibinfo{author}{Bovet, A.}, \bibinfo{author}{Delvenne, J.-C.} \& \bibinfo{author}{Lambiotte, R.}
\newblock \bibinfo{title}{Flow stability for dynamic community detection}.
\newblock \emph{\bibinfo{journal}{Science advances}} \textbf{\bibinfo{volume}{8}}, \bibinfo{pages}{eabj3063} (\bibinfo{year}{2022}).

\bibitem{ghavasieh2024diversity}
\bibinfo{author}{Ghavasieh, A.} \& \bibinfo{author}{De~Domenico, M.}
\newblock \bibinfo{title}{Diversity of information pathways drives sparsity in real-world networks}.
\newblock \emph{\bibinfo{journal}{Nature Physics}} \textbf{\bibinfo{volume}{20}}, \bibinfo{pages}{512--519} (\bibinfo{year}{2024}).

\bibitem{chung2002average}
\bibinfo{author}{Chung, F.} \& \bibinfo{author}{Lu, L.}
\newblock \bibinfo{title}{The average distances in random graphs with given expected degrees}.
\newblock \emph{\bibinfo{journal}{Proceedings of the National Academy of Sciences}} \textbf{\bibinfo{volume}{99}}, \bibinfo{pages}{15879--15882} (\bibinfo{year}{2002}).

\bibitem{boguna2021network}
\bibinfo{author}{Boguna, M.} \emph{et~al.}
\newblock \bibinfo{title}{Network geometry}.
\newblock \emph{\bibinfo{journal}{Nature Reviews Physics}} \textbf{\bibinfo{volume}{3}}, \bibinfo{pages}{114--135} (\bibinfo{year}{2021}).

\bibitem{song2005self}
\bibinfo{author}{Song, C.}, \bibinfo{author}{Havlin, S.} \& \bibinfo{author}{Makse, H.~A.}
\newblock \bibinfo{title}{Self-similarity of complex networks}.
\newblock \emph{\bibinfo{journal}{Nature}} \textbf{\bibinfo{volume}{433}}, \bibinfo{pages}{392--395} (\bibinfo{year}{2005}).

\bibitem{goh2006skeleton}
\bibinfo{author}{Goh, K.-I.}, \bibinfo{author}{Salvi, G.}, \bibinfo{author}{Kahng, B.} \& \bibinfo{author}{Kim, D.}
\newblock \bibinfo{title}{Skeleton and fractal scaling in complex networks}.
\newblock \emph{\bibinfo{journal}{Physical review letters}} \textbf{\bibinfo{volume}{96}}, \bibinfo{pages}{018701} (\bibinfo{year}{2006}).

\bibitem{radicchi2008complex}
\bibinfo{author}{Radicchi, F.}, \bibinfo{author}{Ramasco, J.~J.}, \bibinfo{author}{Barrat, A.} \& \bibinfo{author}{Fortunato, S.}
\newblock \bibinfo{title}{Complex networks renormalization: Flows and fixed points}.
\newblock \emph{\bibinfo{journal}{Physical review letters}} \textbf{\bibinfo{volume}{101}}, \bibinfo{pages}{148701} (\bibinfo{year}{2008}).

\bibitem{rozenfeld2010small}
\bibinfo{author}{Rozenfeld, H.~D.}, \bibinfo{author}{Song, C.} \& \bibinfo{author}{Makse, H.~A.}
\newblock \bibinfo{title}{Small-world to fractal transition in complex networks: a renormalization group approach}.
\newblock \emph{\bibinfo{journal}{Physical review letters}} \textbf{\bibinfo{volume}{104}}, \bibinfo{pages}{025701} (\bibinfo{year}{2010}).

\bibitem{serrano2008self}
\bibinfo{author}{Serrano, M.~{\'A}.}, \bibinfo{author}{Krioukov, D.} \& \bibinfo{author}{Bogun{\'a}, M.}
\newblock \bibinfo{title}{Self-similarity of complex networks and hidden metric spaces}.
\newblock \emph{\bibinfo{journal}{Physical review letters}} \textbf{\bibinfo{volume}{100}}, \bibinfo{pages}{078701} (\bibinfo{year}{2008}).

\bibitem{boguna2009navigability}
\bibinfo{author}{Boguna, M.}, \bibinfo{author}{Krioukov, D.} \& \bibinfo{author}{Claffy, K.~C.}
\newblock \bibinfo{title}{Navigability of complex networks}.
\newblock \emph{\bibinfo{journal}{Nature Physics}} \textbf{\bibinfo{volume}{5}}, \bibinfo{pages}{74--80} (\bibinfo{year}{2009}).

\bibitem{krioukov2010hyperbolic}
\bibinfo{author}{Krioukov, D.}, \bibinfo{author}{Papadopoulos, F.}, \bibinfo{author}{Kitsak, M.}, \bibinfo{author}{Vahdat, A.} \& \bibinfo{author}{Bogun{\'a}, M.}
\newblock \bibinfo{title}{Hyperbolic geometry of complex networks}.
\newblock \emph{\bibinfo{journal}{Physical Review E—Statistical, Nonlinear, and Soft Matter Physics}} \textbf{\bibinfo{volume}{82}}, \bibinfo{pages}{036106} (\bibinfo{year}{2010}).

\bibitem{papadopoulos2012popularity}
\bibinfo{author}{Papadopoulos, F.}, \bibinfo{author}{Kitsak, M.}, \bibinfo{author}{Serrano, M.~{\'A}.}, \bibinfo{author}{Bogun{\'a}, M.} \& \bibinfo{author}{Krioukov, D.}
\newblock \bibinfo{title}{Popularity versus similarity in growing networks}.
\newblock \emph{\bibinfo{journal}{Nature}} \textbf{\bibinfo{volume}{489}}, \bibinfo{pages}{537--540} (\bibinfo{year}{2012}).

\bibitem{barthelemy2011spatial}
\bibinfo{author}{Barth{\'e}lemy, M.}
\newblock \bibinfo{title}{Spatial networks}.
\newblock \emph{\bibinfo{journal}{Physics reports}} \textbf{\bibinfo{volume}{499}}, \bibinfo{pages}{1--101} (\bibinfo{year}{2011}).

\bibitem{goyal2018graph}
\bibinfo{author}{Goyal, P.} \& \bibinfo{author}{Ferrara, E.}
\newblock \bibinfo{title}{Graph embedding techniques, applications, and performance: A survey}.
\newblock \emph{\bibinfo{journal}{Knowledge-Based Systems}} \textbf{\bibinfo{volume}{151}}, \bibinfo{pages}{78--94} (\bibinfo{year}{2018}).

\bibitem{baptista2023zoo}
\bibinfo{author}{Baptista, A.}, \bibinfo{author}{S{\'a}nchez-Garc{\'\i}a, R.~J.}, \bibinfo{author}{Baudot, A.} \& \bibinfo{author}{Bianconi, G.}
\newblock \bibinfo{title}{Zoo guide to network embedding}.
\newblock \emph{\bibinfo{journal}{Journal of Physics: Complexity}} \textbf{\bibinfo{volume}{4}}, \bibinfo{pages}{042001} (\bibinfo{year}{2023}).

\bibitem{brockmann2013hidden}
\bibinfo{author}{Brockmann, D.} \& \bibinfo{author}{Helbing, D.}
\newblock \bibinfo{title}{The hidden geometry of complex, network-driven contagion phenomena}.
\newblock \emph{\bibinfo{journal}{science}} \textbf{\bibinfo{volume}{342}}, \bibinfo{pages}{1337--1342} (\bibinfo{year}{2013}).

\bibitem{iannelli2017effective}
\bibinfo{author}{Iannelli, F.}, \bibinfo{author}{Koher, A.}, \bibinfo{author}{Brockmann, D.}, \bibinfo{author}{H{\"o}vel, P.} \& \bibinfo{author}{Sokolov, I.~M.}
\newblock \bibinfo{title}{Effective distances for epidemics spreading on complex networks}.
\newblock \emph{\bibinfo{journal}{Physical Review E}} \textbf{\bibinfo{volume}{95}}, \bibinfo{pages}{012313} (\bibinfo{year}{2017}).

\bibitem{klamser2024inferring}
\bibinfo{author}{Klamser, P.~P.} \emph{et~al.}
\newblock \bibinfo{title}{Inferring country-specific import risk of diseases from the world air transportation network}.
\newblock \emph{\bibinfo{journal}{PLOS Computational Biology}} \textbf{\bibinfo{volume}{20}}, \bibinfo{pages}{e1011775} (\bibinfo{year}{2024}).

\bibitem{bontorin2023multi}
\bibinfo{author}{Bontorin, S.} \& \bibinfo{author}{De~Domenico, M.}
\newblock \bibinfo{title}{Multi pathways temporal distance unravels the hidden geometry of network-driven processes}.
\newblock \emph{\bibinfo{journal}{Communications Physics}} \textbf{\bibinfo{volume}{6}}, \bibinfo{pages}{129} (\bibinfo{year}{2023}).

\bibitem{coifman2005geometric}
\bibinfo{author}{Coifman, R.~R.} \emph{et~al.}
\newblock \bibinfo{title}{Geometric diffusions as a tool for harmonic analysis and structure definition of data: Diffusion maps}.
\newblock \emph{\bibinfo{journal}{Proceedings of the national academy of sciences}} \textbf{\bibinfo{volume}{102}}, \bibinfo{pages}{7426--7431} (\bibinfo{year}{2005}).

\bibitem{de2017diffusion}
\bibinfo{author}{De~Domenico, M.}
\newblock \bibinfo{title}{Diffusion geometry unravels the emergence of functional clusters in collective phenomena}.
\newblock \emph{\bibinfo{journal}{Physical review letters}} \textbf{\bibinfo{volume}{118}}, \bibinfo{pages}{168301} (\bibinfo{year}{2017}).

\bibitem{barzon2024unraveling}
\bibinfo{author}{Barzon, G.}, \bibinfo{author}{Artime, O.}, \bibinfo{author}{Suweis, S.} \& \bibinfo{author}{Domenico, M.~D.}
\newblock \bibinfo{title}{Unraveling the mesoscale organization induced by network-driven processes}.
\newblock \emph{\bibinfo{journal}{Proceedings of the National Academy of Sciences}} \textbf{\bibinfo{volume}{121}}, \bibinfo{pages}{e2317608121} (\bibinfo{year}{2024}).

\bibitem{estrada2008communicability}
\bibinfo{author}{Estrada, E.} \& \bibinfo{author}{Hatano, N.}
\newblock \bibinfo{title}{Communicability in complex networks}.
\newblock \emph{\bibinfo{journal}{Physical Review E—Statistical, Nonlinear, and Soft Matter Physics}} \textbf{\bibinfo{volume}{77}}, \bibinfo{pages}{036111} (\bibinfo{year}{2008}).

\bibitem{akbarzadeh2018communicability}
\bibinfo{author}{Akbarzadeh, M.} \& \bibinfo{author}{Estrada, E.}
\newblock \bibinfo{title}{Communicability geometry captures traffic flows in cities}.
\newblock \emph{\bibinfo{journal}{Nature human behaviour}} \textbf{\bibinfo{volume}{2}}, \bibinfo{pages}{645--652} (\bibinfo{year}{2018}).

\bibitem{gautreau2008global}
\bibinfo{author}{Gautreau, A.}, \bibinfo{author}{Barrat, A.} \& \bibinfo{author}{Barthelemy, M.}
\newblock \bibinfo{title}{Global disease spread: statistics and estimation of arrival times}.
\newblock \emph{\bibinfo{journal}{Journal of theoretical biology}} \textbf{\bibinfo{volume}{251}}, \bibinfo{pages}{509--522} (\bibinfo{year}{2008}).

\bibitem{belkin2001laplacian}
\bibinfo{author}{Belkin, M.} \& \bibinfo{author}{Niyogi, P.}
\newblock \bibinfo{title}{Laplacian eigenmaps and spectral techniques for embedding and clustering}.
\newblock \emph{\bibinfo{journal}{Advances in neural information processing systems}} \textbf{\bibinfo{volume}{14}} (\bibinfo{year}{2001}).

\bibitem{mohar1991laplacian}
\bibinfo{author}{Mohar, B.}, \bibinfo{author}{Alavi, Y.}, \bibinfo{author}{Chartrand, G.} \& \bibinfo{author}{Oellermann, O.}
\newblock \bibinfo{title}{The laplacian spectrum of graphs}.
\newblock \emph{\bibinfo{journal}{Graph theory, combinatorics, and applications}} \textbf{\bibinfo{volume}{2}}, \bibinfo{pages}{12} (\bibinfo{year}{1991}).

\bibitem{bertagnolli2022functional}
\bibinfo{author}{Bertagnolli, G.} \& \bibinfo{author}{De~Domenico, M.}
\newblock \bibinfo{title}{Functional rich clubs emerging from the diffusion geometry of complex networks}.
\newblock \emph{\bibinfo{journal}{Physical Review Research}} \textbf{\bibinfo{volume}{4}}, \bibinfo{pages}{033185} (\bibinfo{year}{2022}).

\bibitem{bertagnolli2021diffusion}
\bibinfo{author}{Bertagnolli, G.} \& \bibinfo{author}{De~Domenico, M.}
\newblock \bibinfo{title}{Diffusion geometry of multiplex and interdependent systems}.
\newblock \emph{\bibinfo{journal}{Physical Review E}} \textbf{\bibinfo{volume}{103}}, \bibinfo{pages}{042301} (\bibinfo{year}{2021}).

\bibitem{klamser2023enhancing}
\bibinfo{author}{Klamser, P.~P.} \emph{et~al.}
\newblock \bibinfo{title}{Enhancing global preparedness during an ongoing pandemic from partial and noisy data}.
\newblock \emph{\bibinfo{journal}{PNAS nexus}} \textbf{\bibinfo{volume}{2}}, \bibinfo{pages}{pgad192} (\bibinfo{year}{2023}).

\bibitem{kati2024self}
\bibinfo{author}{Kati, Y.}, \bibinfo{author}{Ranft, J.} \& \bibinfo{author}{Lindner, B.}
\newblock \bibinfo{title}{Self-consistent autocorrelation of a disordered kuramoto model in the asynchronous state}.
\newblock \emph{\bibinfo{journal}{Physical Review E}} \textbf{\bibinfo{volume}{110}}, \bibinfo{pages}{054301} (\bibinfo{year}{2024}).

\bibitem{davidsen2024introduction}
\bibinfo{author}{Davidsen, J.}, \bibinfo{author}{Maistrenko, Y.} \& \bibinfo{author}{Showalter, K.}
\newblock \bibinfo{title}{Introduction to focus issue: Chimera states: From theory and experiments to technology and living systems}.
\newblock \emph{\bibinfo{journal}{Chaos: An Interdisciplinary Journal of Nonlinear Science}} \textbf{\bibinfo{volume}{34}} (\bibinfo{year}{2024}).

\bibitem{wang2021effect}
\bibinfo{author}{Wang, Z.} \& \bibinfo{author}{Liu, Z.}
\newblock \bibinfo{title}{Effect of remote signal propagation in an empirical brain network}.
\newblock \emph{\bibinfo{journal}{Chaos: An Interdisciplinary Journal of Nonlinear Science}} \textbf{\bibinfo{volume}{31}} (\bibinfo{year}{2021}).

\bibitem{munoz2010griffiths}
\bibinfo{author}{Munoz, M.~A.}, \bibinfo{author}{Juh{\'a}sz, R.}, \bibinfo{author}{Castellano, C.} \& \bibinfo{author}{{\'O}dor, G.}
\newblock \bibinfo{title}{Griffiths phases on complex networks}.
\newblock \emph{\bibinfo{journal}{Physical review letters}} \textbf{\bibinfo{volume}{105}}, \bibinfo{pages}{128701} (\bibinfo{year}{2010}).

\bibitem{hastings2018transient}
\bibinfo{author}{Hastings, A.} \emph{et~al.}
\newblock \bibinfo{title}{Transient phenomena in ecology}.
\newblock \emph{\bibinfo{journal}{Science}} \textbf{\bibinfo{volume}{361}}, \bibinfo{pages}{eaat6412} (\bibinfo{year}{2018}).

\bibitem{lekien2007lagrangian}
\bibinfo{author}{Lekien, F.}, \bibinfo{author}{Shadden, S.~C.} \& \bibinfo{author}{Marsden, J.~E.}
\newblock \bibinfo{title}{Lagrangian coherent structures in n-dimensional systems}.
\newblock \emph{\bibinfo{journal}{Journal of mathematical physics}} \textbf{\bibinfo{volume}{48}} (\bibinfo{year}{2007}).

\bibitem{haller2015lagrangian}
\bibinfo{author}{Haller, G.}
\newblock \bibinfo{title}{Lagrangian coherent structures}.
\newblock \emph{\bibinfo{journal}{Annual review of fluid mechanics}} \textbf{\bibinfo{volume}{47}}, \bibinfo{pages}{137--162} (\bibinfo{year}{2015}).

\bibitem{haller2011variational}
\bibinfo{author}{Haller, G.}
\newblock \bibinfo{title}{A variational theory of hyperbolic lagrangian coherent structures}.
\newblock \emph{\bibinfo{journal}{Physica D: Nonlinear Phenomena}} \textbf{\bibinfo{volume}{240}}, \bibinfo{pages}{574--598} (\bibinfo{year}{2011}).

\bibitem{amari2016information}
\bibinfo{author}{Amari, S.-i.}
\newblock \emph{\bibinfo{title}{Information geometry and its applications}}, vol. \bibinfo{volume}{194} (\bibinfo{publisher}{Springer}, \bibinfo{year}{2016}).

\bibitem{janke2002information}
\bibinfo{author}{Janke, W.}, \bibinfo{author}{Johnston, D.} \& \bibinfo{author}{Malmini, R.~P.}
\newblock \bibinfo{title}{Information geometry of the ising model on planar random graphs}.
\newblock \emph{\bibinfo{journal}{Physical Review E}} \textbf{\bibinfo{volume}{66}}, \bibinfo{pages}{056119} (\bibinfo{year}{2002}).

\bibitem{kolodrubetz2017geometry}
\bibinfo{author}{Kolodrubetz, M.}, \bibinfo{author}{Sels, D.}, \bibinfo{author}{Mehta, P.} \& \bibinfo{author}{Polkovnikov, A.}
\newblock \bibinfo{title}{Geometry and non-adiabatic response in quantum and classical systems}.
\newblock \emph{\bibinfo{journal}{Physics Reports}} \textbf{\bibinfo{volume}{697}}, \bibinfo{pages}{1--87} (\bibinfo{year}{2017}).

\bibitem{erdmenger2020information}
\bibinfo{author}{Erdmenger, J.}, \bibinfo{author}{Grosvenor, K.} \& \bibinfo{author}{Jefferson, R.}
\newblock \bibinfo{title}{Information geometry in quantum field theory: lessons from simple examples}.
\newblock \emph{\bibinfo{journal}{SciPost Physics}} \textbf{\bibinfo{volume}{8}}, \bibinfo{pages}{073} (\bibinfo{year}{2020}).

\bibitem{floerchinger2023information}
\bibinfo{author}{Floerchinger, S.}
\newblock \bibinfo{title}{Information geometry of euclidean quantum fields}.
\newblock \emph{\bibinfo{journal}{arXiv preprint arXiv:2303.04081}}  (\bibinfo{year}{2023}).

\bibitem{kim2021information}
\bibinfo{author}{Kim, E.-j.}
\newblock \bibinfo{title}{Information geometry, fluctuations, non-equilibrium thermodynamics, and geodesics in complex systems}.
\newblock \emph{\bibinfo{journal}{Entropy}} \textbf{\bibinfo{volume}{23}}, \bibinfo{pages}{1393} (\bibinfo{year}{2021}).

\bibitem{ito2018stochastic}
\bibinfo{author}{Ito, S.}
\newblock \bibinfo{title}{Stochastic thermodynamic interpretation of information geometry}.
\newblock \emph{\bibinfo{journal}{Physical review letters}} \textbf{\bibinfo{volume}{121}}, \bibinfo{pages}{030605} (\bibinfo{year}{2018}).

\bibitem{janke2004information}
\bibinfo{author}{Janke, W.}, \bibinfo{author}{Johnston, D.} \& \bibinfo{author}{Kenna, R.}
\newblock \bibinfo{title}{Information geometry and phase transitions}.
\newblock \emph{\bibinfo{journal}{Physica A: Statistical Mechanics and its Applications}} \textbf{\bibinfo{volume}{336}}, \bibinfo{pages}{181--186} (\bibinfo{year}{2004}).

\bibitem{carollo2020geometry}
\bibinfo{author}{Carollo, A.}, \bibinfo{author}{Valenti, D.} \& \bibinfo{author}{Spagnolo, B.}
\newblock \bibinfo{title}{Geometry of quantum phase transitions}.
\newblock \emph{\bibinfo{journal}{Physics Reports}} \textbf{\bibinfo{volume}{838}}, \bibinfo{pages}{1--72} (\bibinfo{year}{2020}).

\bibitem{banchi2014quantum}
\bibinfo{author}{Banchi, L.}, \bibinfo{author}{Giorda, P.} \& \bibinfo{author}{Zanardi, P.}
\newblock \bibinfo{title}{Quantum information-geometry of dissipative quantum phase transitions}.
\newblock \emph{\bibinfo{journal}{Physical Review E}} \textbf{\bibinfo{volume}{89}}, \bibinfo{pages}{022102} (\bibinfo{year}{2014}).

\bibitem{alexandrov2023information}
\bibinfo{author}{Alexandrov, A.} \& \bibinfo{author}{Gorsky, A.}
\newblock \bibinfo{title}{Information geometry and synchronization phase transition in the kuramoto model}.
\newblock \emph{\bibinfo{journal}{Physical Review E}} \textbf{\bibinfo{volume}{107}}, \bibinfo{pages}{044211} (\bibinfo{year}{2023}).

\bibitem{metz2025dynamical}
\bibinfo{author}{Metz, F.~L.}
\newblock \bibinfo{title}{Dynamical mean-field theory of complex systems on sparse directed networks}.
\newblock \emph{\bibinfo{journal}{Physical Review Letters}} \textbf{\bibinfo{volume}{134}}, \bibinfo{pages}{037401} (\bibinfo{year}{2025}).

\end{thebibliography}

\end{document}